\documentclass[amsmath,a4paper,aps,twocolumn,prl,reprint,superscriptaddress]{revtex4-1}

\usepackage{bm}
\usepackage{graphicx} 
\usepackage{color}

\begin{document}
 
\title{Theory of long-range ultracold atom-molecule photoassociation}

\author{Jes\'{u}s P\'{e}rez-R\'{i}os 
\footnote{Electronic mail: jperezri@purdue.edu}}

\affiliation{Department of Physics and Astronomy, Purdue University, 
West Lafayette, IN 47907, USA}

\affiliation{Laboratoire Aim\'{e} Cotton, CNRS/ Universit\'e Paris-Sud/ ENS Cachan, B\^{a}t 505, 
91405 Orsay, France}

\author{Maxence Lepers}

\affiliation{Laboratoire Aim\'{e} Cotton, CNRS/ Universit\'e Paris-Sud/ ENS Cachan, B\^{a}t 505, 
91405 Orsay, France}

\author{Olivier Dulieu}

\affiliation{Laboratoire Aim\'{e} Cotton, CNRS/ Universit\'e Paris-Sud/ ENS Cachan, B\^{a}t 505, 
91405 Orsay, France}

\date{\today}

\begin{abstract}
 
The creation of ultracold molecules is currently limited to diatomic species. In this letter we present a theoretical description of the photoassociation of ultracold atoms and molecules to create ultracold excited triatomic molecules, thus being a novel example of light-assisted ultracold chemical reaction. The calculation of the photoassociation rate of ultracold Cs atoms with ultracold Cs$_2$ molecules in their rovibrational ground state is reported, based on the solution of the quantum dynamics involving the atom-molecule long-range interactions, and assuming a model potential for the short-range physics. The rate for the formation of excited Cs$_3$ molecules is predicted to be comparable with currently observed atom-atom photoassociation rates. We formulate an experimental proposal to observe this process relying on the available techniques of optical lattices and standard photoassociation spectroscopy.
\end{abstract} 


\maketitle

Ultracold dilute gases at temperatures much lower than 1~mK offer new opportunities for the study of elementary reactive processes between atoms and molecules. As emphasized in seminal review articles \cite{krems2008,bell2009} this so-called ultracold chemistry is freed from averaging on large velocity distributions. Thus in this regime the quantum nature of the processes can be accessed, like the presence of resonances, or the sensitivity to long-range interactions which can induce anisotropic arrangements prior to the reaction. One of the most advanced experiments in this direction is undoubtedly under progress in JILA at Boulder, where quantum threshold collisions induced by tunneling through a centrifugal barrier between fermionic ultracold KRb molecules have been detected \cite{ni2010,demiranda2011}. Many experiments demonstrated that conditions suitable for inducing chemical reactions between ultracold ground state alkali-metal atoms and molecules tightly bound in their electronic ground state are reachable. Ultracold collisions between Cs atoms and Cs$_2$ \cite{zahzam2006, staanum2006} or LiCs \cite{deiglmayr2011} molecules, of Rb and Cs atoms with RbCs molecules \cite{hudson2008}, or of K and Rb atoms with KRb molecules \cite{ospelkaus2010a} have been detected through atom trap losses. Three-body recombination in a Rb quantum gas has been characterized \cite{haerter2013}, and features associated to universal $N$-body (up to $N=5$) resonances induced by collisions between atoms and weakly-bound molecules have been observed in quantum degenerate gases \cite{kraemer2006,ferlaino2009,zaccanti2009,pollack2009,zenesini2013,zenesini2014}.

However the deep understanding of the collisional dynamics is still to come as none of these achievements were able to characterize the nature and the state distribution of the final products \cite{nesbitt2012,gonzalez-martinez2014}. The heavy mass of alkali-metal species leads to a huge density of resonant states related to the numerous rovibrational levels accessible during an atom/molecule or a molecule/molecule collision, preventing them to be individually characterized. Models connecting the standard treatment of long-range interactions between particles to a statistical description of the resonances mostly resulting from short-range couplings have been developed \cite{mayle2012,mayle2013,croft2014,gonzalez-martinez2014}. The main quantitative uncertainty of such models arises from the estimation of the actual amount of resonant states which are indeed active during the collision.

In this letter, we propose to use the well-known photoassociation (PA) process to get more insight into such collisions between ultracold atoms and molecules. PA of an ultracold atom pair is a laser-assisted collision creating an electronically-excited diatomic molecule often in a weakly-bound energy level \cite{thorsheim1987,jones2006}. Under favorable circumstances such a short-lived molecule can decay by spontaneous emission (SE) into a rovibrational level of a stable electronic state. The PA+SE process played a crucial role in creating samples of stable ultracold molecules \cite{fioretti1998,dulieu2009,carr2009,bruzewicz2014}. This is the simplest example of the formation of a chemical bond under ultracold conditions. The PA step is mostly controlled by long-range atom-atom interactions \cite{pillet1997,cote1998,bohn1999}, while the SE step involves short-range interaction \cite{fioretti1998,dion2001}. The knowledge of atom-atom interactions is thus required to fully describe the PA+SE process, which is possible in most cases.

Here we discuss the PA of ultracold alkali-metal atoms and diatomic molecules to create ultracold trimers. We evaluate the rate for PA of ultracold ground state Cs atoms and ultracold ground-state Cs$_2$ molecules in their lowest rovibrational level, revealing a magnitude similar to the rates observed for atom-atom PA. Our model only involves the long-range interactions between the particles, thus neglecting the influence of the short-range interactions responsible for a possible high density of resonant states. Thus the problem is formally similar to the PA of a pair of atoms of different species. We propose an experimental way to observe the excited Cs$^{*}_{3}$ molecules using an nearly-degenerate quantum gas of Cs atoms and molecules trapped in an optical lattice \cite{danzl2010}. By comparison to the present simplified model, such an observation would shed light on the link between short-range and long-range physics in ultracold processes.

We consider the interaction between two charge distributions A and B with total angular momentum $\vec{J}_A$ and $\vec{J}_B$ and separated by a distance $R$ along an axis $Z$ joining their center-of-mass and oriented from the molecule toward the atom. It can be split according to two distance ranges: (i) the long-range domain where interactions are determined by the individual properties of A and B and can be accurately calculated, and (ii) the short-range domain where complex chemical interactions take place. Both domains connect around the LeRoy radius $R_{LR}$ accounting for the size of each distribution \cite{leroy1974}. For the rest of the paper we consider a Cs($6^{2}$S$_{1/2}$) atom and a Cs$_2$ molecule in the lowest vibrational ($v=0$) and rotational level $N=0$ of its electronic ground state X$^{1}\Sigma_{g}^{+}$ as the initial state $|i\rangle$ of PA. We have $R^i_{LR}=42a_0$ ($a_0=0.052917721092$~nm). The laser frequency is assumed to be slightly detuned by $\delta_{PA}$ from the D1 or D2 Cs line, so that PA populates energy levels of the Cs$^{*}_{3}$ complex located below the Cs($6^{2}$P$_{3/2}$)+Cs$_2$(X$^{1}\Sigma_{g}^{+}$,$v=0$,$N=0$) dissociation limit. For these final states we have $R^f_{LR}=46a_0$. We ignore the Cs($6^{2}$P$_{j}$) hyperfine structure for simplicity, and we will consider experimental conditions where it can be a secure approximation.

Following our previous work, the long-range excited atom-ground-state molecule interaction is treated within the second-order degenerate perturbation theory \cite{lepers2011b,lepers2011c}. In brief, the Hamiltonian of the system is written as $\hat{H}=\hat{H}_{0}+\hat{W}$, where $\hat{H}_{0}$ refers to the energy of the individual particles at infinity. The atomic states are labeled as $|l=1,s=1/2,j,\omega\rangle$, where $j=1/2,3/2$ is the total angular momentum quantum number and $\omega$ its projection on the $Z$-axis. The molecular states $|\mathrm{X}^{1}\Sigma_{g}^{+},v=0,N,m_N\rangle$ involves the projection $m_N$ of $\vec{N}$ onto $Z$. The mutual rotation of the atom and the molecule is not introduced yet, so that the total projection quantum number $\Omega=m_N+\omega$ characterizes the trimer states which will be referred to as $|j,\omega,N,m_N;\Omega;p\rangle$, where $p=(-1)^{N}$ is the parity. The $\hat{W}(R)$ operator includes the first-order quadrupole-quadrupole term ($\hat{V}_{qq} (R)\propto 1/R^{5}$, if $N \ne 0$) \cite{lepers2010}, and the second order dipole-dipole interaction ($\hat{V}_{dd}^{(2)}(R)\propto 1/R^{6}$) \cite{lepers2011a}.

\begin{figure}[t]
\includegraphics[width=8cm,angle=0.]{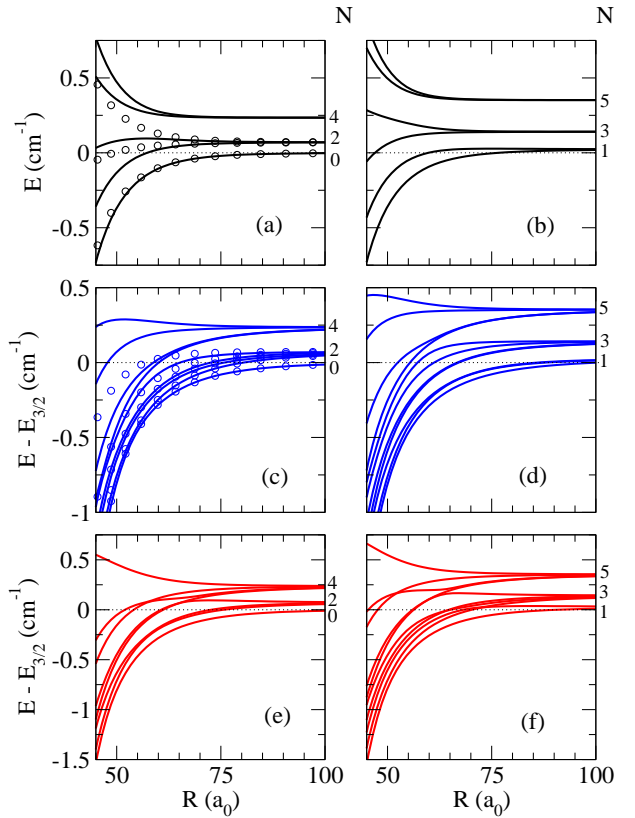}
\caption[]{(Color online) Long-range potential energy curves between a ground-state Cs$_{2}$(X$^{1}\Sigma_{g}^{+}$,$v=0$,$N$) molecule and an excited atom Cs(6$^{2}$P$_{j}$) as a function of the scattering coordinate $R$. (a),(b): $j=1/2, \Omega=1/2$; (c), (d): $j=3/2, \Omega=1/2$; (e), (f): $j=3/2, \Omega=3/2$. The origin of energies is taken at the Cs$_{2}$(X,$\nu_{d}=0$,$N$=0)+Cs(6$^{2}$P$_{j=1/2}$) limit. Closed circles in (a) and (c) correspond to PECs computed with only $N=0, 2$ levels in the Hamiltonian.} 
\label{fig:pecs}
\end{figure}

The Hamiltonian written in the $|j,\omega,N,m_N;\Omega\rangle$ basis is diagonalized, yielding the adiabatic potential energy curves (PECs) for large atom-molecule distances displayed in Fig.~\ref{fig:pecs}, for $j=1/2,3/2$, $\Omega=1/2,3/2$, and including rotational levels up to $N=5$. The complex structure of the PECs is due to the coupling between the pure electrostatic interaction and the rotation of the molecule. In particular, these PECs cannot be expressed anymore as a pure $R^{-n}$ expansion. For instance the lowest curve labeled with $N=0$ at large distance for $j=3/2, \Omega$ (panels c and e of Fig.~\ref{fig:pecs}) is more attractive -- and thus more favorable for long-range PA-- due to the coupling with the $N=2$ state (see Fig.~2 in Ref.~\cite{lepers2011b}). We also computed the PECs including only $N=0, 2$ levels in the Hamiltonian, showing that upper rotational levels only slightly affect the lowest PECs for $N=0$. At $R \approx R_{LR}$ the PECs are attractive by at most 1~cm$^{-1}$, so that we limit our study to PA detunings smaller than 1~cm$^{-1}$ below the Cs D2 line. Note that the PECs correlated to the Cs(6$^{2}$P$_{1/2}$)+Cs$_{2}$(X$^{1}\Sigma_{g}^{+}$,$v=0$,$N$) are obviously simpler than in the $j=3/2$ case, which could yield preferable conditions for a first experimental implementation. 

For $R < R_{LR}$, each diagonal term of the above Hamiltonian is smoothly matched to a Lennard-Jones potential $V_{LJ}=D_{LJ}(C_{LJ}/R^6)((C_{LJ}/R^6)-1)$. The $C_{LJ}$ factor is chosen equal to the computed $C_6$ coefficient \cite{lepers2011a}, while the depth $D_{LJ}$ is taken large enough to ensure that the short-range part of the wavefunctions have a negligible contribution to the ultracold dynamics. For the initial PA state we have $C^i_{6} = 12101$~a.u. and $D^i_{e} = 10867.2$~cm$^{-1}$. For the final PA states of interest here, namely with $j=3/2$, $N = 0, 2$ and $|\Omega|=1/2$, the parameters are $C^f_{6}(N=0) = 60311$~a.u., $D^f_{e}(N=0) = 1500$~cm$^{-1}$, and $C^f_{6}(N=2) = 25311$~a.u., $D^f_{e}(N=2) = 500$~cm$^{-1}$.

The radial wavefunctions $|\Psi_{f}(\Omega,v')\rangle$ associated to the eigenvalue $E_f(v')$ of the coupled states described by the above PECs are calculated with the mapped Fourier grid Hamiltonian (MFGH) method \cite{kokoouline1999} suitable for calculations at large distances as the grid step is mapped onto the local kinetic energy of the channels. We used a grid of $N_g=4145$ points between $4a_{0}$ and $7000a_{0}$ with the MFGH scaling parameter of  $\beta = 0.11$ \cite{kokoouline1999} to ensure a proper convergence of the $E_f(v')$ values.

\begin{figure}[h]
\includegraphics[width=0.5\textwidth]{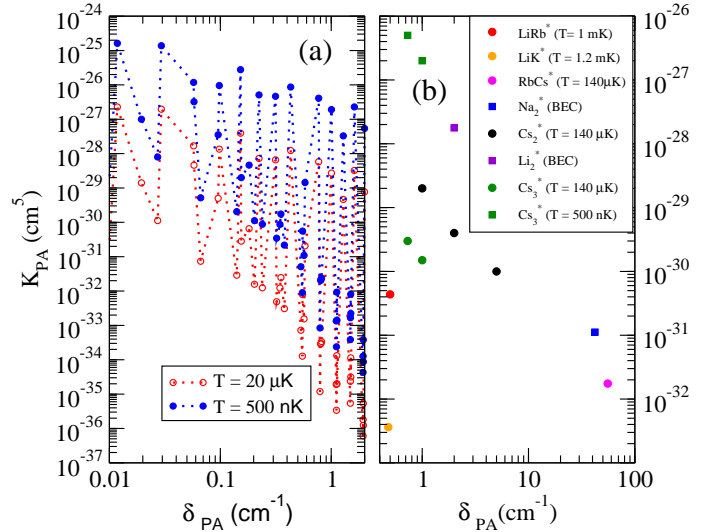}
\caption[]{(Color online) (a) Normalized PA rate $K_{\rm{PA}}$ as a function of the detuning $\delta_{\rm{PA}}$ of the PA laser  with respect to Cs$_{2}$(X$^{1}\Sigma_{g}^{+}$,$v=0$,$N = 0$) + Cs($6^{2}$P$_{3/2}$). The points correspond to vibrational levels $v'$ of Cs$^*_3$ for $|\Omega|= 1/2$. The rates are calculated at $T=20\, \mu$K (open red circles) and $T=500$~nK (closed blue circles). (b) Comparison of computed $K_{PA}$ values at 140~$\mu$K and 500~nK for two Cs$^*_3$vibrational levels $v'$ at 0.770 and 1.008~cm$^{-1}$ with the theoretical value in Cs$_2$ \cite{drag2000}, and several experimental rates for LiK \cite{ridinger2011}, LiRb \cite{dutta2014}, RbCs \cite{kerman2004a}, Na$_2$ \cite{mckenzie2002}, and Li$_2$ \cite{prodan2003}.}
\label{fig:nparate}
\end{figure}

We assume that PA takes place during a $s$-wave atom-molecule collision at an energy $E_i=k_BT$, where $k_B$ is the Boltzman constant and $T$ the related temperature. The associated continuum radial wavefunction $\Psi_{i}$ is computed by solving the Schr\"odinger equation for the entrance channel with the Numerov method. In the perturbative regime, if the PA laser with intensity $I_{\rm{PA}}$, detuning $\delta_{\rm{PA}}$ and polarization $\vec{\epsilon}_{\rm{PA}}$, hits a level $v'$ of the Cs$^*_3$ complex, the PA rate is expressed as~\cite{pillet1997}
\begin{equation}
\label{eq:parate}
R_{\rm{PA}}\left(v',T\right) =  A_{if}\left( \frac{3}{2\pi} \right)^{3/2} 
\frac{h}{2}n_{mol}\frac{I_{\rm{PA}}}{I_{0}}\Lambda_{T}^{3}\Omega_{if}^{2}|\langle\Psi_{i}|\Psi_{f}(\Omega,v')\rangle|^{2},
\end{equation}

\noindent where $\Lambda_{T}=h\sqrt{1/3\mu k_{B}T}$, $\mu$ is the atom-molecule reduced mass, $n_{mol}$ stands for the Cs$_2$ density, and $A_{if}(\vec{\epsilon}_{PA})$ is an angular factor depending on $\vec{\epsilon}_{\rm{PA}}$. Its value strongly depends on the initial state of the particles, and we took $A_{if}=1/2$ as a typical example. As the detuning is assumed to be small, the $|i\rangle \rightarrow |f\rangle$ transition dipole moment is taken as the one for the $6^2S_{1/2} \rightarrow 6^2P_{3/2}$ transition, leading to the corresponding atomic Rabi frequency $\Omega_{if}$ and saturation intensity $I_{0}=1.1$~mW/cm$^{2}$ \cite{pillet1997}. For the purpose of comparison among various experimental implementations, it is useful to define the normalized PA rate $K_{\rm{PA}}(T)=R_{\rm{PA}}(T)/{n_{mol}/\phi_{\rm{PA}}}$ (conveniently expressed in cm$^5$) where $\phi_{\rm{PA}}=I_{\rm{PA}}\lambda_{\rm{PA}}/hc$ is the PA laser photon flux at the wavelength $\lambda_{\rm{PA}}$ \cite{cote1998,drag2000}. The results for $K_{\rm{PA}}(T)$ are displayed in Fig.~\ref{fig:nparate}a for two different temperatures:  at $T=20\,\mu$K typical for a magneto-optical trap (MOT),  and at $T=500$~nK representative of a gas close to the quantum degeneracy (QD) regime. Like in the case of heteronuclear diatomic molecules \cite{azizi2004}, the oscillating pattern reflects the strong variations of the spatial overlap between the initial and final radial wavefunctions with the detuning. As expected from Eq.~(\ref{eq:parate}), PA close to the QD regime and for small detunings exhibits a large rate, as already demonstrated in the experiments on Li$_2$ \cite{prodan2003} and Na$_2$ \cite{mckenzie2002}. We mention that the coupling of the long-range multipolar interactions with the rotational energy quoted above also contributes to enhance the PA rate as the entrance channel $N=0$ is more attractive. This first principal result is further illustrated in Fig.~\ref{fig:nparate}b showing that the predicted PA rate in a MOT environment is comparable to the rates observed with various heteronuclear diatomic species. We recall that the PA rate for Cs$_2$ at 140~$\mu$K \cite{drag2000} was larger than most of the PA rates observed under MOT conditions, as it was enhanced by the presence of a long-range double-well in the excited PEC.  The large magnitude of the atom/molecule PA rate under QD conditions shows that such an experiment is achievable just like its Li-Li counterpart of Ref~\cite{prodan2003}.
 
Although the results look very promising, we must discuss whether the conditions of density and temperature for atoms and molecules needed for the formation of Cs$^{*}_{3}$ can be satisfied with the available experimental techniques. A density $n_{mol}$ of Cs$_{2}$(X$^{1}$$\Sigma_{g}^{+}$,$\nu$=0,$N$=0) molecules of about $\approx 10^{10}$ to $10^{12}$~cm$^{-3}$ and temperature $T\sim 500$~nK, can be obtained departing from an ultracold sample of ground-state Cs atoms~\cite{danzl2010} with similar density. These are favorable conditions to initiate a PA experiment where products would be indirectly detected through the loss of molecules. 

\begin{figure}[t]
\includegraphics[width=0.4\textwidth,angle=0.]{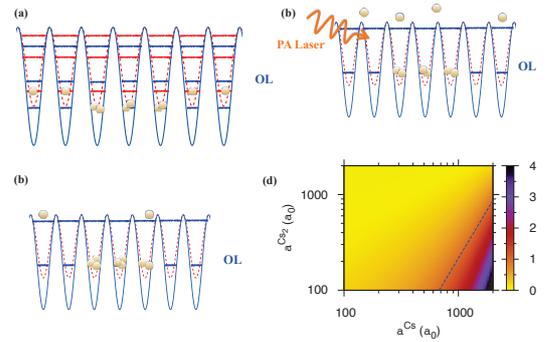}
\caption[]{(Color online) Scheme of the experimental proposal. (a) Preparation of a MI state for both atoms and molecules. (b) The PA laser is applied to a mixed molecular MI -- atomic SF state. (c) Excited Cs$_{3}^{*}$ molecules are created, and trapped after radiative decay into stable Cs$_3$ trimers. (d) Contour plot for the ratio of critical OL intensities for Cs and Cs$_2$ (Eq.~(\ref{eq:ratio})) as a function of their respective scattering lengths. The dashed line indicates where this ratio equals unity.}
\label{fig:expprop}
\end{figure}

As discussed above, the direct detection of products is also a central objective to better understand the full dynamics of the process, which could benefit from the presence of an optical lattice (OL). The scheme of our experimental proposal is presented in Fig.~\ref{fig:expprop}, based on the conditions of Ref.~\cite{danzl2010}. The motional degrees of freedom of the atoms are controlled by an OL with wavelength $\lambda_{OL}$=1064.5 nm. The OL intensity is tuned in order to induce a Mott-insulator (MI) state with preferentially two atoms per lattice site (see {e.g.} Ref.~\cite{bloch2008}). The magnetoassociation technique is then applied, creating ground-state Cs$_{2}$ molecules in a weakly-bound vibrational level. This population is transferred to the Cs$_2$ rovibrational ground state by means of stimulated Raman adiabatic passage (STIRAP) technique which yields a molecular MI state to a good approximation~\cite{danzl2010} (Fig.~\ref{fig:expprop}a).

During this sequence a fraction of ultracold atoms remains unpaired. We propose to tune the OL intensity to drive the transition from a MI phase to a superfluid (SF) phase for atoms, keeping the molecules in the MI state (Fig.~\ref{fig:expprop}b). Assuming a deep optical lattice with spacing between two successive sites $d=\lambda_{OL}/2$, the MI-to-SF phase transition for the species $\beta$ ($\beta \equiv$~Cs or Cs$_2$) occurs when the lattice depth $V_{0}^{\beta} \propto \alpha_d^{\beta} I_{\rm{OL}}$ (where $I_{\rm{OL}}$ is the corresponding intensity and $\alpha_d^{\beta}$ the dynamic dipole polarizability of the species at $\lambda_{OL}$) satisfies the critical value given by~\cite{bloch2008}
\begin{equation}
\label{eq:OLdepth}
\left(\frac{(V_{0}^{\beta})_c}{E_{R}^{\beta}}\right)=\frac{1}{4}\ln^2\left( \frac{\sqrt{2}d}{\pi a^{\beta}} \left(\frac{U^{\beta}}{J^{\beta}}\right)_{c}\right)
\end{equation}
\noindent
where $E_{R}^{\beta}$ is the recoil energy of the considered species and $a^{\beta}$ is the corresponding scattering length. The critical value $(U^{\beta}/J^{\beta})_{c}$ of the ratio between the on-site interaction $U^{\beta}$ and the tunneling energy $J^{\beta}$ depends on the number of $\beta$ particles per OL sites $\bar{n}$. We will assume $\bar{n}$=1 hereafter, leading to $(U/J)_{c}=29.36$~\cite{bloch2008} in the case of a simple cubic lattice. At $\lambda_{OL}=1064.5$~nm, the recoil energies are such that $E_R^{\rm{Cs}}/k_B=64$~nK and $E_{R}^{\rm{Cs}_{2}}/k_B=32$~nK. Equation~(\ref{eq:OLdepth}) shows that the critical value depends on the scattering length $a^{\alpha}$ of the two species (taken as positive quantities as a MI phase can be achieved) but only through a logarithmic way. While the variations of $a^{\rm{Cs}}$ are well established \cite{chin2004b,berninger2013}, those of $a^{\rm{Cs_2}}$ are totally unknown. A reasonable assumption is to consider that the experiment is performed under conditions where $a^{\rm{Cs}}$ and $a^{\rm{Cs_2}}$ are not taking vanishing or infinite values, such that $d/a^{\alpha} \gg 1$. This leads to the estimation of the ratio of the critical OL intensities for each species 
\begin{equation}
\frac{(I_{\rm{OL}}^{\rm{Cs_2}})_c}{(I_{\rm{OL}}^{\rm{Cs}})_c} \approx \frac{\alpha_d^{\rm{Cs}}E_{R}^{\rm{Cs}_2}}{\alpha_d^{\rm{Cs}_2}E_{R}^{\rm{Cs}}} = \frac{\alpha_d^{\rm{Cs}}}{2\alpha_d^{\rm{Cs}_2}},
\label{eq:ratio}
\end{equation}
\noindent which must be smaller than unity to ensure that the MI--SF transition is carried out on the atoms without releasing the MI phase of the molecules. Assuming that $\alpha_d^{\rm{Cs_2}} \approx 2\alpha_d^{\rm{Cs}}$ \cite{vexiau2011}, the ratio indeed amounts to about 1/4. This estimate is further confirmed in Fig.~\ref{fig:expprop}d, showing that $(I_{\rm{OL}}^{\rm{Cs_2}})_c/(I_{\rm{OL}}^{\rm{Cs}})_c$ is indeed smaller than unity for a broad range of scattering lengths (Fig.~\ref{fig:expprop}c) for both species.

The PA laser is not expected to significantly disturb such an hybrid system. Indeed, for $\lambda_{\rm{PA}} \approx 852$~nm (or $\sim 11737$~cm$^{-1}$, close to the D2 Cs line) the Cs$_2$ ground state molecule will only weakly absorb such a light due to unfavorable Franck-Condon factors with excited molecular states, hence the related scattering force will be negligible (see Fig.~3 of Ref.~\cite{vexiau2011}). 

Finally, assuming that the photoassociated Cs$_3^*$ complex has a radiative lifetime similar to the one of electronically-excited Cs or Cs$_2$ species, the products of the decay could be discriminated in the experiment (Fig.~\ref{fig:expprop}c). On one hand, the decay of Cs$_3^*$ back into three Cs atoms or a Cs/Cs$_2$ pair with a kinetic energy release will result in a loss of the fragments from the OL. On the other hand, if stable Cs$_3$ molecules are created, they will most probably be trapped in the OL, as their recoil energy is even smaller than the Cs$_2$ one as their larger dipole polarizability induces a deeper trap. A branching ratio between the two decay channels could thus be inferred from the benefit of the "`half-collision"' character of the process leading to bound products.
 
The observation of ultracold trimers in a PA experiment will shed light on the role of short-range interactions in ultracold collisions in between two extreme situations: either PA results into well identified resonant features assigned to Cs$^*_3$ levels determined by pure long-range interactions (as in the present model), or PA will reveal a quasicontinuum of levels induced by the strong coupling of such long-range states with numerous resonances of the trimer (as in the approach of Refs~\cite{mayle2012,mayle2013}. The presented approach can be generalized to stable ultracold Cs$_2$ molecules in their lowest $^{3}\Sigma_u^+$ state, but this case could be quite challenging as the PA wavelength will reach the region of numerous Cs$_2$ excited levels (see Fig.~6 in Ref.~\cite{vexiau2011}). Other species can also be investigated, as for instance the association of Li atoms and K$_2$ ground state molecules where the K$_2$ rotational energy is similar to the Li$^*$ spin-orbit splitting, inducing complex patterns in the long-range PECs \cite{lepers2011c}.

The authors acknowledge enlightening discussions with H.-C. N\"{a}gerl and K. Lauber in relation with the experimental proposal, and W. Stwalley for helpful discussions about the PA possibility of $^{3}\Sigma_u^+$ molecules and ultracold atoms. JPR acknowledge funding from IFRAF during the development of this project.  


\begin{thebibliography}{49}
\expandafter\ifx\csname natexlab\endcsname\relax\def\natexlab#1{#1}\fi
\expandafter\ifx\csname bibnamefont\endcsname\relax
  \def\bibnamefont#1{#1}\fi
\expandafter\ifx\csname bibfnamefont\endcsname\relax
  \def\bibfnamefont#1{#1}\fi
\expandafter\ifx\csname citenamefont\endcsname\relax
  \def\citenamefont#1{#1}\fi
\expandafter\ifx\csname url\endcsname\relax
  \def\url#1{\texttt{#1}}\fi
\expandafter\ifx\csname urlprefix\endcsname\relax\def\urlprefix{URL }\fi
\providecommand{\bibinfo}[2]{#2}
\providecommand{\eprint}[2][]{\url{#2}}

\bibitem[{\citenamefont{Krems}(2008)}]{krems2008}
\bibinfo{author}{\bibfnamefont{R.~V.} \bibnamefont{Krems}},
  \bibinfo{journal}{Phys. Chem. Chem. Phys.} \textbf{\bibinfo{volume}{10}},
  \bibinfo{pages}{4079} (\bibinfo{year}{2008}).

\bibitem[{\citenamefont{Bell and Softley}(20099)}]{bell2009}
\bibinfo{author}{\bibfnamefont{M.}~\bibnamefont{Bell}} \bibnamefont{and}
  \bibinfo{author}{\bibfnamefont{T.~P.} \bibnamefont{Softley}},
  \bibinfo{journal}{Mol. Phys.} \textbf{\bibinfo{volume}{107}},
  \bibinfo{pages}{99} (\bibinfo{year}{20099}).

\bibitem[{\citenamefont{Ni et~al.}(2010)\citenamefont{Ni, Ospelkaus, Wang,
  Qu\'em\'ener, Neyenhuis, de~Miranda, Bohn, Ye, and Jin}}]{ni2010}
\bibinfo{author}{\bibfnamefont{K.-K.} \bibnamefont{Ni}},
  \bibinfo{author}{\bibfnamefont{S.}~\bibnamefont{Ospelkaus}},
  \bibinfo{author}{\bibfnamefont{D.}~\bibnamefont{Wang}},
  \bibinfo{author}{\bibfnamefont{G.}~\bibnamefont{Qu\'em\'ener}},
  \bibinfo{author}{\bibfnamefont{B.}~\bibnamefont{Neyenhuis}},
  \bibinfo{author}{\bibfnamefont{M.~H.~G.} \bibnamefont{de~Miranda}},
  \bibinfo{author}{\bibfnamefont{J.~L.} \bibnamefont{Bohn}},
  \bibinfo{author}{\bibfnamefont{J.}~\bibnamefont{Ye}}, \bibnamefont{and}
  \bibinfo{author}{\bibfnamefont{D.~S.} \bibnamefont{Jin}},
  \bibinfo{journal}{Nature} \textbf{\bibinfo{volume}{464}},
  \bibinfo{pages}{1324} (\bibinfo{year}{2010}).

\bibitem[{\citenamefont{de~Miranda et~al.}(2011)\citenamefont{de~Miranda,
  Chotia, Neyenhuis, Wang, Qu{\'e}m{\'e}ner, Ospelkaus, Bohn, Ye, and
  Jin}}]{demiranda2011}
\bibinfo{author}{\bibfnamefont{M.}~\bibnamefont{de~Miranda}},
  \bibinfo{author}{\bibfnamefont{A.}~\bibnamefont{Chotia}},
  \bibinfo{author}{\bibfnamefont{B.}~\bibnamefont{Neyenhuis}},
  \bibinfo{author}{\bibfnamefont{D.}~\bibnamefont{Wang}},
  \bibinfo{author}{\bibfnamefont{G.}~\bibnamefont{Qu{\'e}m{\'e}ner}},
  \bibinfo{author}{\bibfnamefont{S.}~\bibnamefont{Ospelkaus}},
  \bibinfo{author}{\bibfnamefont{J.}~\bibnamefont{Bohn}},
  \bibinfo{author}{\bibfnamefont{J.}~\bibnamefont{Ye}}, \bibnamefont{and}
  \bibinfo{author}{\bibfnamefont{D.}~\bibnamefont{Jin}}, \bibinfo{journal}{Nat.
  Phys.} \textbf{\bibinfo{volume}{7}}, \bibinfo{pages}{502}
  (\bibinfo{year}{2011}).

\bibitem[{\citenamefont{Zahzam et~al.}(2006)\citenamefont{Zahzam, Vogt,
  Mudrich, Comparat, and Pillet}}]{zahzam2006}
\bibinfo{author}{\bibfnamefont{N.}~\bibnamefont{Zahzam}},
  \bibinfo{author}{\bibfnamefont{T.}~\bibnamefont{Vogt}},
  \bibinfo{author}{\bibfnamefont{M.}~\bibnamefont{Mudrich}},
  \bibinfo{author}{\bibfnamefont{D.}~\bibnamefont{Comparat}}, \bibnamefont{and}
  \bibinfo{author}{\bibfnamefont{P.}~\bibnamefont{Pillet}},
  \bibinfo{journal}{Phys. Rev. Lett.} \textbf{\bibinfo{volume}{96}},
  \bibinfo{pages}{023202} (\bibinfo{year}{2006}).

\bibitem[{\citenamefont{Staanum et~al.}(2006)\citenamefont{Staanum, Kraft,
  Lange, Wester, and Weidem\"{u}ller}}]{staanum2006}
\bibinfo{author}{\bibfnamefont{P.}~\bibnamefont{Staanum}},
  \bibinfo{author}{\bibfnamefont{S.~D.} \bibnamefont{Kraft}},
  \bibinfo{author}{\bibfnamefont{J.}~\bibnamefont{Lange}},
  \bibinfo{author}{\bibfnamefont{R.}~\bibnamefont{Wester}}, \bibnamefont{and}
  \bibinfo{author}{\bibfnamefont{M.}~\bibnamefont{Weidem\"{u}ller}},
  \bibinfo{journal}{Phys. Rev. Lett.} \textbf{\bibinfo{volume}{96}},
  \bibinfo{pages}{023201} (\bibinfo{year}{2006}).

\bibitem[{\citenamefont{Deiglmayr et~al.}(2011)\citenamefont{Deiglmayr, Repp,
  Wester, Dulieu, and Weidem\"uller}}]{deiglmayr2011}
\bibinfo{author}{\bibfnamefont{J.}~\bibnamefont{Deiglmayr}},
  \bibinfo{author}{\bibfnamefont{M.}~\bibnamefont{Repp}},
  \bibinfo{author}{\bibfnamefont{R.}~\bibnamefont{Wester}},
  \bibinfo{author}{\bibfnamefont{O.}~\bibnamefont{Dulieu}}, \bibnamefont{and}
  \bibinfo{author}{\bibfnamefont{M.}~\bibnamefont{Weidem\"uller}},
  \bibinfo{journal}{Phys. Chem. Chem. Phys.} \textbf{\bibinfo{volume}{13}},
  \bibinfo{pages}{19101} (\bibinfo{year}{2011}).

\bibitem[{\citenamefont{Hudson et~al.}(2008)\citenamefont{Hudson, Gilfoy,
  Kotochigova, Sage, and DeMille}}]{hudson2008}
\bibinfo{author}{\bibfnamefont{E.~R.} \bibnamefont{Hudson}},
  \bibinfo{author}{\bibfnamefont{N.~B.} \bibnamefont{Gilfoy}},
  \bibinfo{author}{\bibfnamefont{S.}~\bibnamefont{Kotochigova}},
  \bibinfo{author}{\bibfnamefont{J.~M.} \bibnamefont{Sage}}, \bibnamefont{and}
  \bibinfo{author}{\bibfnamefont{D.}~\bibnamefont{DeMille}},
  \bibinfo{journal}{Phys. Rev. Lett.} \textbf{\bibinfo{volume}{100}},
  \bibinfo{pages}{203201} (\bibinfo{year}{2008}).

\bibitem[{\citenamefont{Ospelkaus et~al.}(2010)\citenamefont{Ospelkaus, Ni,
  Wang, de~Miranda, Neyenhuis, Qu\'em\'ener, Julienne, Bohn, Jin, and
  Ye}}]{ospelkaus2010a}
\bibinfo{author}{\bibfnamefont{S.}~\bibnamefont{Ospelkaus}},
  \bibinfo{author}{\bibfnamefont{K.-K.} \bibnamefont{Ni}},
  \bibinfo{author}{\bibfnamefont{D.}~\bibnamefont{Wang}},
  \bibinfo{author}{\bibfnamefont{M.~H.~G.} \bibnamefont{de~Miranda}},
  \bibinfo{author}{\bibfnamefont{B.}~\bibnamefont{Neyenhuis}},
  \bibinfo{author}{\bibfnamefont{G.}~\bibnamefont{Qu\'em\'ener}},
  \bibinfo{author}{\bibfnamefont{P.~S.} \bibnamefont{Julienne}},
  \bibinfo{author}{\bibfnamefont{J.}~\bibnamefont{Bohn}},
  \bibinfo{author}{\bibfnamefont{D.~S.} \bibnamefont{Jin}}, \bibnamefont{and}
  \bibinfo{author}{\bibfnamefont{J.}~\bibnamefont{Ye}},
  \bibinfo{journal}{Science} \textbf{\bibinfo{volume}{327}},
  \bibinfo{pages}{853} (\bibinfo{year}{2010}).

\bibitem[{\citenamefont{H\"arter et~al.}(2013)\citenamefont{H\"arter, Kr\"ukow,
  Dei{\ss}, Drews, Tiemann, and Hecker~Denschlag}}]{haerter2013}
\bibinfo{author}{\bibfnamefont{A.}~\bibnamefont{H\"arter}},
  \bibinfo{author}{\bibfnamefont{A.}~\bibnamefont{Kr\"ukow}},
  \bibinfo{author}{\bibfnamefont{M.}~\bibnamefont{Dei{\ss}}},
  \bibinfo{author}{\bibfnamefont{B.}~\bibnamefont{Drews}},
  \bibinfo{author}{\bibfnamefont{E.}~\bibnamefont{Tiemann}}, \bibnamefont{and}
  \bibinfo{author}{\bibfnamefont{J.}~\bibnamefont{Hecker~Denschlag}},
  \bibinfo{journal}{Nature Phys.} \textbf{\bibinfo{volume}{9}},
  \bibinfo{pages}{512} (\bibinfo{year}{2013}).

\bibitem[{\citenamefont{Kraemer et~al.}(2006)\citenamefont{Kraemer, Mark,
  Waldburger, Danzl, Chin, Engeser, Lange, Pilch, Jaakkola, N\"{a}gerl
  et~al.}}]{kraemer2006}
\bibinfo{author}{\bibfnamefont{T.}~\bibnamefont{Kraemer}},
  \bibinfo{author}{\bibfnamefont{M.}~\bibnamefont{Mark}},
  \bibinfo{author}{\bibfnamefont{P.}~\bibnamefont{Waldburger}},
  \bibinfo{author}{\bibfnamefont{J.~G.} \bibnamefont{Danzl}},
  \bibinfo{author}{\bibfnamefont{C.}~\bibnamefont{Chin}},
  \bibinfo{author}{\bibfnamefont{B.}~\bibnamefont{Engeser}},
  \bibinfo{author}{\bibfnamefont{A.~D.} \bibnamefont{Lange}},
  \bibinfo{author}{\bibfnamefont{K.}~\bibnamefont{Pilch}},
  \bibinfo{author}{\bibfnamefont{A.}~\bibnamefont{Jaakkola}},
  \bibinfo{author}{\bibfnamefont{H.-C.} \bibnamefont{N\"{a}gerl}},
  \bibnamefont{et~al.}, \bibinfo{journal}{Nature}
  \textbf{\bibinfo{volume}{440}}, \bibinfo{pages}{315} (\bibinfo{year}{2006}).

\bibitem[{\citenamefont{Ferlaino et~al.}(2009)\citenamefont{Ferlaino, Knoop,
  Berninger, Harm, D'Incao, N\"agerl, and Grimm}}]{ferlaino2009}
\bibinfo{author}{\bibfnamefont{F.}~\bibnamefont{Ferlaino}},
  \bibinfo{author}{\bibfnamefont{S.}~\bibnamefont{Knoop}},
  \bibinfo{author}{\bibfnamefont{M.}~\bibnamefont{Berninger}},
  \bibinfo{author}{\bibfnamefont{W.}~\bibnamefont{Harm}},
  \bibinfo{author}{\bibfnamefont{J.~P.} \bibnamefont{D'Incao}},
  \bibinfo{author}{\bibfnamefont{H.-C.} \bibnamefont{N\"agerl}},
  \bibnamefont{and} \bibinfo{author}{\bibfnamefont{R.}~\bibnamefont{Grimm}},
  \bibinfo{journal}{Phys. Rev. Lett.} \textbf{\bibinfo{volume}{102}},
  \bibinfo{pages}{140401} (\bibinfo{year}{2009}).

\bibitem[{\citenamefont{Zaccanti et~al.}(2009)\citenamefont{Zaccanti, Deissler,
  DӅrrico, Fattori, Jona-Lasinio, M\"uller, Roati, Inguscio, and
  Modugno}}]{zaccanti2009}
\bibinfo{author}{\bibfnamefont{M.}~\bibnamefont{Zaccanti}},
  \bibinfo{author}{\bibfnamefont{B.}~\bibnamefont{Deissler}},
  \bibinfo{author}{\bibfnamefont{C.}~\bibnamefont{DӅrrico}},
  \bibinfo{author}{\bibfnamefont{M.}~\bibnamefont{Fattori}},
  \bibinfo{author}{\bibfnamefont{M.}~\bibnamefont{Jona-Lasinio}},
  \bibinfo{author}{\bibfnamefont{S.}~\bibnamefont{M\"uller}},
  \bibinfo{author}{\bibfnamefont{G.}~\bibnamefont{Roati}},
  \bibinfo{author}{\bibfnamefont{M.}~\bibnamefont{Inguscio}}, \bibnamefont{and}
  \bibinfo{author}{\bibfnamefont{G.}~\bibnamefont{Modugno}},
  \bibinfo{journal}{Nature Phys.} \textbf{\bibinfo{volume}{5}},
  \bibinfo{pages}{586} (\bibinfo{year}{2009}).

\bibitem[{\citenamefont{Pollack et~al.}(2009)\citenamefont{Pollack, Dries, and
  Hulet}}]{pollack2009}
\bibinfo{author}{\bibfnamefont{S.~E.} \bibnamefont{Pollack}},
  \bibinfo{author}{\bibfnamefont{D.}~\bibnamefont{Dries}}, \bibnamefont{and}
  \bibinfo{author}{\bibfnamefont{R.~G.} \bibnamefont{Hulet}},
  \bibinfo{journal}{Science} \textbf{\bibinfo{volume}{326}},
  \bibinfo{pages}{1683} (\bibinfo{year}{2009}).

\bibitem[{\citenamefont{Zenesini et~al.}(2013)\citenamefont{Zenesini, Huang,
  Berninger, Besler, N\"agerl, Ferlaino, Grimm, Greene, and von
  Stecher}}]{zenesini2013}
\bibinfo{author}{\bibfnamefont{A.}~\bibnamefont{Zenesini}},
  \bibinfo{author}{\bibfnamefont{B.}~\bibnamefont{Huang}},
  \bibinfo{author}{\bibfnamefont{M.}~\bibnamefont{Berninger}},
  \bibinfo{author}{\bibfnamefont{S.}~\bibnamefont{Besler}},
  \bibinfo{author}{\bibfnamefont{H.-C.} \bibnamefont{N\"agerl}},
  \bibinfo{author}{\bibfnamefont{F.}~\bibnamefont{Ferlaino}},
  \bibinfo{author}{\bibfnamefont{R.}~\bibnamefont{Grimm}},
  \bibinfo{author}{\bibfnamefont{C.~H.} \bibnamefont{Greene}},
  \bibnamefont{and} \bibinfo{author}{\bibfnamefont{J.}~\bibnamefont{von
  Stecher}}, \bibinfo{journal}{New J. Phys.} \textbf{\bibinfo{volume}{15}},
  \bibinfo{pages}{043040} (\bibinfo{year}{2013}).

\bibitem[{\citenamefont{Zenesini et~al.}(2014)\citenamefont{Zenesini, Huang,
  Berninger, N\"agerl, Ferlaino, and Grimm}}]{zenesini2014}
\bibinfo{author}{\bibfnamefont{A.}~\bibnamefont{Zenesini}},
  \bibinfo{author}{\bibfnamefont{B.}~\bibnamefont{Huang}},
  \bibinfo{author}{\bibfnamefont{M.}~\bibnamefont{Berninger}},
  \bibinfo{author}{\bibfnamefont{H.-C.} \bibnamefont{N\"agerl}},
  \bibinfo{author}{\bibfnamefont{F.}~\bibnamefont{Ferlaino}}, \bibnamefont{and}
  \bibinfo{author}{\bibfnamefont{R.}~\bibnamefont{Grimm}},
  \bibinfo{journal}{Phys. Rev. A} \textbf{\bibinfo{volume}{90}},
  \bibinfo{pages}{022704} (\bibinfo{year}{2014}).

\bibitem[{\citenamefont{Nesbitt}(2012)}]{nesbitt2012}
\bibinfo{author}{\bibfnamefont{D.~J.} \bibnamefont{Nesbitt}},
  \bibinfo{journal}{Chem. Rev.} \textbf{\bibinfo{volume}{112}},
  \bibinfo{pages}{5062} (\bibinfo{year}{2012}).

\bibitem[{\citenamefont{Gonz{\'a}lez-Mart{\'\i}nez
  et~al.}(2014)\citenamefont{Gonz{\'a}lez-Mart{\'\i}nez, Dulieu,
  Larr{\'e}garay, and Bonnet}}]{gonzalez-martinez2014}
\bibinfo{author}{\bibfnamefont{M.}~\bibnamefont{Gonz{\'a}lez-Mart{\'\i}nez}},
  \bibinfo{author}{\bibfnamefont{O.}~\bibnamefont{Dulieu}},
  \bibinfo{author}{\bibfnamefont{P.}~\bibnamefont{Larr{\'e}garay}},
  \bibnamefont{and} \bibinfo{author}{\bibfnamefont{L.}~\bibnamefont{Bonnet}},
  \bibinfo{journal}{Phys. Rev. A} \textbf{\bibinfo{volume}{90}},
  \bibinfo{pages}{052716} (\bibinfo{year}{2014}).

\bibitem[{\citenamefont{Mayle et~al.}(2012)\citenamefont{Mayle, Ruzic, and
  Bohn}}]{mayle2012}
\bibinfo{author}{\bibfnamefont{M.}~\bibnamefont{Mayle}},
  \bibinfo{author}{\bibfnamefont{B.~P.} \bibnamefont{Ruzic}}, \bibnamefont{and}
  \bibinfo{author}{\bibfnamefont{J.~L.} \bibnamefont{Bohn}},
  \bibinfo{journal}{Phys. Rev. A} \textbf{\bibinfo{volume}{85}},
  \bibinfo{pages}{062712} (\bibinfo{year}{2012}).

\bibitem[{\citenamefont{Mayle et~al.}(2013)\citenamefont{Mayle, Qu\'em\'ener,
  Ruzic, and Bohn}}]{mayle2013}
\bibinfo{author}{\bibfnamefont{M.}~\bibnamefont{Mayle}},
  \bibinfo{author}{\bibfnamefont{G.}~\bibnamefont{Qu\'em\'ener}},
  \bibinfo{author}{\bibfnamefont{B.~P.} \bibnamefont{Ruzic}}, \bibnamefont{and}
  \bibinfo{author}{\bibfnamefont{J.~L.} \bibnamefont{Bohn}},
  \bibinfo{journal}{Phys. Rev. A} \textbf{\bibinfo{volume}{87}},
  \bibinfo{pages}{012709} (\bibinfo{year}{2013}).

\bibitem[{\citenamefont{Croft and Bohn}(2014)}]{croft2014}
\bibinfo{author}{\bibfnamefont{J.~F.~E.} \bibnamefont{Croft}} \bibnamefont{and}
  \bibinfo{author}{\bibfnamefont{J.~L.} \bibnamefont{Bohn}},
  \bibinfo{journal}{Phys. Rev. A} \textbf{\bibinfo{volume}{89}},
  \bibinfo{pages}{012714} (\bibinfo{year}{2014}).

\bibitem[{\citenamefont{Thorsheim et~al.}(1987)\citenamefont{Thorsheim, Weiner,
  and Julienne}}]{thorsheim1987}
\bibinfo{author}{\bibfnamefont{H.~R.} \bibnamefont{Thorsheim}},
  \bibinfo{author}{\bibfnamefont{J.}~\bibnamefont{Weiner}}, \bibnamefont{and}
  \bibinfo{author}{\bibfnamefont{P.~S.} \bibnamefont{Julienne}},
  \bibinfo{journal}{Phys. Rev. Lett.} \textbf{\bibinfo{volume}{58}},
  \bibinfo{pages}{2420} (\bibinfo{year}{1987}).

\bibitem[{\citenamefont{Jones et~al.}(2006)\citenamefont{Jones, Tiesinga, Lett,
  and Julienne}}]{jones2006}
\bibinfo{author}{\bibfnamefont{K.~M.} \bibnamefont{Jones}},
  \bibinfo{author}{\bibfnamefont{E.}~\bibnamefont{Tiesinga}},
  \bibinfo{author}{\bibfnamefont{P.~D.} \bibnamefont{Lett}}, \bibnamefont{and}
  \bibinfo{author}{\bibfnamefont{P.~S.} \bibnamefont{Julienne}},
  \bibinfo{journal}{Rev. Mod. Phys.} \textbf{\bibinfo{volume}{78}},
  \bibinfo{pages}{483} (\bibinfo{year}{2006}).

\bibitem[{\citenamefont{Fioretti et~al.}(1998)\citenamefont{Fioretti, Comparat,
  Crubellier, Dulieu, Masnou-Seeuws, and Pillet}}]{fioretti1998}
\bibinfo{author}{\bibfnamefont{A.}~\bibnamefont{Fioretti}},
  \bibinfo{author}{\bibfnamefont{D.}~\bibnamefont{Comparat}},
  \bibinfo{author}{\bibfnamefont{A.}~\bibnamefont{Crubellier}},
  \bibinfo{author}{\bibfnamefont{O.}~\bibnamefont{Dulieu}},
  \bibinfo{author}{\bibfnamefont{F.}~\bibnamefont{Masnou-Seeuws}},
  \bibnamefont{and} \bibinfo{author}{\bibfnamefont{P.}~\bibnamefont{Pillet}},
  \bibinfo{journal}{Phys. Rev. Lett.} \textbf{\bibinfo{volume}{80}},
  \bibinfo{pages}{4402} (\bibinfo{year}{1998}).

\bibitem[{\citenamefont{Dulieu and Gabbanini}(2009)}]{dulieu2009}
\bibinfo{author}{\bibfnamefont{O.}~\bibnamefont{Dulieu}} \bibnamefont{and}
  \bibinfo{author}{\bibfnamefont{C.}~\bibnamefont{Gabbanini}},
  \bibinfo{journal}{Rep. Prog. Phys.} \textbf{\bibinfo{volume}{72}},
  \bibinfo{pages}{086401} (\bibinfo{year}{2009}).

\bibitem[{\citenamefont{Carr and Ye}(2009)}]{carr2009}
\bibinfo{author}{\bibfnamefont{L.~D.} \bibnamefont{Carr}} \bibnamefont{and}
  \bibinfo{author}{\bibfnamefont{J.}~\bibnamefont{Ye}}, \bibinfo{journal}{New
  J. Phys.} \textbf{\bibinfo{volume}{11}}, \bibinfo{pages}{055009}
  (\bibinfo{year}{2009}).

\bibitem[{\citenamefont{Bruzewicz et~al.}(2014)\citenamefont{Bruzewicz,
  Gustavsson, Shimasaki, and DeMille}}]{bruzewicz2014}
\bibinfo{author}{\bibfnamefont{C.~D.} \bibnamefont{Bruzewicz}},
  \bibinfo{author}{\bibfnamefont{M.}~\bibnamefont{Gustavsson}},
  \bibinfo{author}{\bibfnamefont{T.}~\bibnamefont{Shimasaki}},
  \bibnamefont{and} \bibinfo{author}{\bibfnamefont{D.}~\bibnamefont{DeMille}},
  \bibinfo{journal}{New J. Phys.} \textbf{\bibinfo{volume}{16}},
  \bibinfo{pages}{023018} (\bibinfo{year}{2014}).

\bibitem[{\citenamefont{Pillet et~al.}(1997)\citenamefont{Pillet, Crubellier,
  Bleton, Dulieu, Nosbaum, Mourachko, and Masnou-Seeuws}}]{pillet1997}
\bibinfo{author}{\bibfnamefont{P.}~\bibnamefont{Pillet}},
  \bibinfo{author}{\bibfnamefont{A.}~\bibnamefont{Crubellier}},
  \bibinfo{author}{\bibfnamefont{A.}~\bibnamefont{Bleton}},
  \bibinfo{author}{\bibfnamefont{O.}~\bibnamefont{Dulieu}},
  \bibinfo{author}{\bibfnamefont{P.}~\bibnamefont{Nosbaum}},
  \bibinfo{author}{\bibfnamefont{I.}~\bibnamefont{Mourachko}},
  \bibnamefont{and}
  \bibinfo{author}{\bibfnamefont{F.}~\bibnamefont{Masnou-Seeuws}},
  \bibinfo{journal}{J. Phys. B} \textbf{\bibinfo{volume}{30}},
  \bibinfo{pages}{2801} (\bibinfo{year}{1997}).

\bibitem[{\citenamefont{C\^ot\'e and Dalgarno}(1998)}]{cote1998}
\bibinfo{author}{\bibfnamefont{R.}~\bibnamefont{C\^ot\'e}} \bibnamefont{and}
  \bibinfo{author}{\bibfnamefont{A.}~\bibnamefont{Dalgarno}},
  \bibinfo{journal}{Phys. Rev. A} \textbf{\bibinfo{volume}{58}},
  \bibinfo{pages}{498} (\bibinfo{year}{1998}).

\bibitem[{\citenamefont{Bohn and Julienne}(1999)}]{bohn1999}
\bibinfo{author}{\bibfnamefont{J.~L.} \bibnamefont{Bohn}} \bibnamefont{and}
  \bibinfo{author}{\bibfnamefont{P.~S.} \bibnamefont{Julienne}},
  \bibinfo{journal}{Phys. Rev. A} \textbf{\bibinfo{volume}{60}},
  \bibinfo{pages}{414} (\bibinfo{year}{1999}).

\bibitem[{\citenamefont{Dion et~al.}(2001)\citenamefont{Dion, Drag, Dulieu,
  Tolra, Masnou-Seeuws, and Pillet}}]{dion2001}
\bibinfo{author}{\bibfnamefont{C.~M.} \bibnamefont{Dion}},
  \bibinfo{author}{\bibfnamefont{C.}~\bibnamefont{Drag}},
  \bibinfo{author}{\bibfnamefont{O.}~\bibnamefont{Dulieu}},
  \bibinfo{author}{\bibfnamefont{B.~L.} \bibnamefont{Tolra}},
  \bibinfo{author}{\bibfnamefont{F.}~\bibnamefont{Masnou-Seeuws}},
  \bibnamefont{and} \bibinfo{author}{\bibfnamefont{P.}~\bibnamefont{Pillet}},
  \bibinfo{journal}{Phys. Rev. Lett.} \textbf{\bibinfo{volume}{86}},
  \bibinfo{pages}{2253} (\bibinfo{year}{2001}).

\bibitem[{\citenamefont{Danzl et~al.}(2010)\citenamefont{Danzl, Mark, Haller,
  Gustavsson, Hart, Aldegunde, Hutson, and N\"agerl}}]{danzl2010}
\bibinfo{author}{\bibfnamefont{J.~G.} \bibnamefont{Danzl}},
  \bibinfo{author}{\bibfnamefont{M.~J.} \bibnamefont{Mark}},
  \bibinfo{author}{\bibfnamefont{E.}~\bibnamefont{Haller}},
  \bibinfo{author}{\bibfnamefont{M.}~\bibnamefont{Gustavsson}},
  \bibinfo{author}{\bibfnamefont{R.}~\bibnamefont{Hart}},
  \bibinfo{author}{\bibfnamefont{J.}~\bibnamefont{Aldegunde}},
  \bibinfo{author}{\bibfnamefont{J.~M.} \bibnamefont{Hutson}},
  \bibnamefont{and} \bibinfo{author}{\bibfnamefont{H.-C.}
  \bibnamefont{N\"agerl}}, \bibinfo{journal}{Nature Phys.}
  \textbf{\bibinfo{volume}{6}}, \bibinfo{pages}{265} (\bibinfo{year}{2010}).

\bibitem[{\citenamefont{{Le Roy}}(1974)}]{leroy1974}
\bibinfo{author}{\bibfnamefont{R.~J.} \bibnamefont{{Le Roy}}},
  \bibinfo{journal}{Can. J. Phys.} \textbf{\bibinfo{volume}{52}},
  \bibinfo{pages}{246} (\bibinfo{year}{1974}).

\bibitem[{\citenamefont{Lepers and Dulieu}(2011{\natexlab{a}})}]{lepers2011b}
\bibinfo{author}{\bibfnamefont{M.}~\bibnamefont{Lepers}} \bibnamefont{and}
  \bibinfo{author}{\bibfnamefont{O.}~\bibnamefont{Dulieu}},
  \bibinfo{journal}{Europhys. J. D} \textbf{\bibinfo{volume}{65}},
  \bibinfo{pages}{113} (\bibinfo{year}{2011}{\natexlab{a}}).

\bibitem[{\citenamefont{Lepers and Dulieu}(2011{\natexlab{b}})}]{lepers2011c}
\bibinfo{author}{\bibfnamefont{M.}~\bibnamefont{Lepers}} \bibnamefont{and}
  \bibinfo{author}{\bibfnamefont{O.}~\bibnamefont{Dulieu}},
  \bibinfo{journal}{Phys. Chem. Chem. Phys.} \textbf{\bibinfo{volume}{13}},
  \bibinfo{pages}{19106} (\bibinfo{year}{2011}{\natexlab{b}}).

\bibitem[{\citenamefont{Lepers et~al.}(2010)\citenamefont{Lepers, Dulieu, and
  Kokoouline}}]{lepers2010}
\bibinfo{author}{\bibfnamefont{M.}~\bibnamefont{Lepers}},
  \bibinfo{author}{\bibfnamefont{O.}~\bibnamefont{Dulieu}}, \bibnamefont{and}
  \bibinfo{author}{\bibfnamefont{V.}~\bibnamefont{Kokoouline}},
  \bibinfo{journal}{Phys. Rev. A} \textbf{\bibinfo{volume}{82}},
  \bibinfo{pages}{042711} (\bibinfo{year}{2010}).

\bibitem[{\citenamefont{Lepers et~al.}(2011)\citenamefont{Lepers, Vexiau,
  Bouloufa, Dulieu, and Kokoouline}}]{lepers2011a}
\bibinfo{author}{\bibfnamefont{M.}~\bibnamefont{Lepers}},
  \bibinfo{author}{\bibfnamefont{R.}~\bibnamefont{Vexiau}},
  \bibinfo{author}{\bibfnamefont{N.}~\bibnamefont{Bouloufa}},
  \bibinfo{author}{\bibfnamefont{O.}~\bibnamefont{Dulieu}}, \bibnamefont{and}
  \bibinfo{author}{\bibfnamefont{V.}~\bibnamefont{Kokoouline}},
  \bibinfo{journal}{Phys. Rev. A} \textbf{\bibinfo{volume}{83}},
  \bibinfo{pages}{042707} (\bibinfo{year}{2011}).

\bibitem[{\citenamefont{Kokoouline et~al.}(1999)\citenamefont{Kokoouline,
  Dulieu, Kosloff, and Masnou-Seeuws}}]{kokoouline1999}
\bibinfo{author}{\bibfnamefont{V.}~\bibnamefont{Kokoouline}},
  \bibinfo{author}{\bibfnamefont{O.}~\bibnamefont{Dulieu}},
  \bibinfo{author}{\bibfnamefont{R.}~\bibnamefont{Kosloff}}, \bibnamefont{and}
  \bibinfo{author}{\bibfnamefont{F.}~\bibnamefont{Masnou-Seeuws}},
  \bibinfo{journal}{J. Chem. Phys.} \textbf{\bibinfo{volume}{110}},
  \bibinfo{pages}{9865} (\bibinfo{year}{1999}).

\bibitem[{\citenamefont{Drag et~al.}(2000)\citenamefont{Drag, Tolra, Dulieu,
  Comparat, Vatasescu, Boussen, Guibal, Crubellier, and Pillet}}]{drag2000}
\bibinfo{author}{\bibfnamefont{C.}~\bibnamefont{Drag}},
  \bibinfo{author}{\bibfnamefont{B.~L.} \bibnamefont{Tolra}},
  \bibinfo{author}{\bibfnamefont{O.}~\bibnamefont{Dulieu}},
  \bibinfo{author}{\bibfnamefont{D.}~\bibnamefont{Comparat}},
  \bibinfo{author}{\bibfnamefont{M.}~\bibnamefont{Vatasescu}},
  \bibinfo{author}{\bibfnamefont{S.}~\bibnamefont{Boussen}},
  \bibinfo{author}{\bibfnamefont{S.}~\bibnamefont{Guibal}},
  \bibinfo{author}{\bibfnamefont{A.}~\bibnamefont{Crubellier}},
  \bibnamefont{and} \bibinfo{author}{\bibfnamefont{P.}~\bibnamefont{Pillet}},
  \bibinfo{journal}{IEEE J. Quant. Electron.} \textbf{\bibinfo{volume}{36}},
  \bibinfo{pages}{1378} (\bibinfo{year}{2000}).

\bibitem[{\citenamefont{Azizi et~al.}(2004)\citenamefont{Azizi, Aymar, and
  Dulieu}}]{azizi2004}
\bibinfo{author}{\bibfnamefont{S.}~\bibnamefont{Azizi}},
  \bibinfo{author}{\bibfnamefont{M.}~\bibnamefont{Aymar}}, \bibnamefont{and}
  \bibinfo{author}{\bibfnamefont{O.}~\bibnamefont{Dulieu}},
  \bibinfo{journal}{Eur. Phys. J. D} \textbf{\bibinfo{volume}{31}},
  \bibinfo{pages}{195} (\bibinfo{year}{2004}).

\bibitem[{\citenamefont{Prodan et~al.}(2003)\citenamefont{Prodan, Pichler,
  Junker, Hulet, and Bohn}}]{prodan2003}
\bibinfo{author}{\bibfnamefont{I.~D.} \bibnamefont{Prodan}},
  \bibinfo{author}{\bibfnamefont{M.}~\bibnamefont{Pichler}},
  \bibinfo{author}{\bibfnamefont{M.}~\bibnamefont{Junker}},
  \bibinfo{author}{\bibfnamefont{R.~G.} \bibnamefont{Hulet}}, \bibnamefont{and}
  \bibinfo{author}{\bibfnamefont{J.~L.} \bibnamefont{Bohn}},
  \bibinfo{journal}{Phys. Rev. Lett.} \textbf{\bibinfo{volume}{91}},
  \bibinfo{pages}{080402} (\bibinfo{year}{2003}).

\bibitem[{\citenamefont{McKenzie et~al.}(2002)\citenamefont{McKenzie,
  Denschlag, H\"{a}ffner, Browaeys, de~Araujo, Fatemi, Jones, Simsaran, Cho,
  Simoni et~al.}}]{mckenzie2002}
\bibinfo{author}{\bibfnamefont{C.}~\bibnamefont{McKenzie}},
  \bibinfo{author}{\bibfnamefont{J.~H.} \bibnamefont{Denschlag}},
  \bibinfo{author}{\bibfnamefont{H.}~\bibnamefont{H\"{a}ffner}},
  \bibinfo{author}{\bibfnamefont{A.}~\bibnamefont{Browaeys}},
  \bibinfo{author}{\bibfnamefont{L.~E.} \bibnamefont{de~Araujo}},
  \bibinfo{author}{\bibfnamefont{F.}~\bibnamefont{Fatemi}},
  \bibinfo{author}{\bibfnamefont{K.~M.} \bibnamefont{Jones}},
  \bibinfo{author}{\bibfnamefont{J.}~\bibnamefont{Simsaran}},
  \bibinfo{author}{\bibfnamefont{D.}~\bibnamefont{Cho}},
  \bibinfo{author}{\bibfnamefont{A.}~\bibnamefont{Simoni}},
  \bibnamefont{et~al.}, \bibinfo{journal}{Phys. Rev. Lett.}
  \textbf{\bibinfo{volume}{88}}, \bibinfo{pages}{120403}
  (\bibinfo{year}{2002}).

\bibitem[{\citenamefont{Ridinger et~al.}(2011)\citenamefont{Ridinger,
  Chaudhuri, Salez, Fernandez, Bouloufa, Dulieu, Salomon, and
  Chevy}}]{ridinger2011}
\bibinfo{author}{\bibfnamefont{A.}~\bibnamefont{Ridinger}},
  \bibinfo{author}{\bibfnamefont{S.}~\bibnamefont{Chaudhuri}},
  \bibinfo{author}{\bibfnamefont{T.}~\bibnamefont{Salez}},
  \bibinfo{author}{\bibfnamefont{D.~R.} \bibnamefont{Fernandez}},
  \bibinfo{author}{\bibfnamefont{N.}~\bibnamefont{Bouloufa}},
  \bibinfo{author}{\bibfnamefont{O.}~\bibnamefont{Dulieu}},
  \bibinfo{author}{\bibfnamefont{C.}~\bibnamefont{Salomon}}, \bibnamefont{and}
  \bibinfo{author}{\bibfnamefont{F.}~\bibnamefont{Chevy}},
  \bibinfo{journal}{Europhys. Lett.} \textbf{\bibinfo{volume}{96}},
  \bibinfo{pages}{33001} (\bibinfo{year}{2011}).

\bibitem[{\citenamefont{Dutta et~al.}(2014)\citenamefont{Dutta, Lorenz, Altaf,
  Elliott, and Chen}}]{dutta2014}
\bibinfo{author}{\bibfnamefont{S.}~\bibnamefont{Dutta}},
  \bibinfo{author}{\bibfnamefont{J.}~\bibnamefont{Lorenz}},
  \bibinfo{author}{\bibfnamefont{A.}~\bibnamefont{Altaf}},
  \bibinfo{author}{\bibfnamefont{D.~S.} \bibnamefont{Elliott}},
  \bibnamefont{and} \bibinfo{author}{\bibfnamefont{Y.~P.} \bibnamefont{Chen}},
  \bibinfo{journal}{Phys. Rev. A} \textbf{\bibinfo{volume}{89}},
  \bibinfo{pages}{020702} (\bibinfo{year}{2014}).

\bibitem[{\citenamefont{Kerman et~al.}(2004)\citenamefont{Kerman, Sage, Sainis,
  Bergeman, and DeMille}}]{kerman2004a}
\bibinfo{author}{\bibfnamefont{A.~J.} \bibnamefont{Kerman}},
  \bibinfo{author}{\bibfnamefont{J.~M.} \bibnamefont{Sage}},
  \bibinfo{author}{\bibfnamefont{S.}~\bibnamefont{Sainis}},
  \bibinfo{author}{\bibfnamefont{T.}~\bibnamefont{Bergeman}}, \bibnamefont{and}
  \bibinfo{author}{\bibfnamefont{D.}~\bibnamefont{DeMille}},
  \bibinfo{journal}{Phys. Rev. Lett.} \textbf{\bibinfo{volume}{92}},
  \bibinfo{pages}{033004} (\bibinfo{year}{2004}).

\bibitem[{\citenamefont{Bloch et~al.}(2008)\citenamefont{Bloch, Dalibard, and
  Zwerger}}]{bloch2008}
\bibinfo{author}{\bibfnamefont{I.}~\bibnamefont{Bloch}},
  \bibinfo{author}{\bibfnamefont{J.}~\bibnamefont{Dalibard}}, \bibnamefont{and}
  \bibinfo{author}{\bibfnamefont{W.}~\bibnamefont{Zwerger}},
  \bibinfo{journal}{Rev. Mod. Phys.} \textbf{\bibinfo{volume}{80}},
  \bibinfo{pages}{885} (\bibinfo{year}{2008}).

\bibitem[{\citenamefont{Chin et~al.}(2004)\citenamefont{Chin, Vuleti\'c,
  Kerman, Chu, Tiesinga, Leo, and Williams}}]{chin2004b}
\bibinfo{author}{\bibfnamefont{C.}~\bibnamefont{Chin}},
  \bibinfo{author}{\bibfnamefont{V.}~\bibnamefont{Vuleti\'c}},
  \bibinfo{author}{\bibfnamefont{A.~J.} \bibnamefont{Kerman}},
  \bibinfo{author}{\bibfnamefont{S.}~\bibnamefont{Chu}},
  \bibinfo{author}{\bibfnamefont{E.}~\bibnamefont{Tiesinga}},
  \bibinfo{author}{\bibfnamefont{P.~J.} \bibnamefont{Leo}}, \bibnamefont{and}
  \bibinfo{author}{\bibfnamefont{C.~J.} \bibnamefont{Williams}},
  \bibinfo{journal}{Phys. Rev. A} \textbf{\bibinfo{volume}{70}},
  \bibinfo{pages}{032701} (\bibinfo{year}{2004}).

\bibitem[{\citenamefont{Berninger et~al.}(2013)\citenamefont{Berninger,
  Zenesini, Huang, Harm, N\"agerl, Ferlaino, Grimm, Julienne, and
  Hutson}}]{berninger2013}
\bibinfo{author}{\bibfnamefont{M.}~\bibnamefont{Berninger}},
  \bibinfo{author}{\bibfnamefont{A.}~\bibnamefont{Zenesini}},
  \bibinfo{author}{\bibfnamefont{B.}~\bibnamefont{Huang}},
  \bibinfo{author}{\bibfnamefont{W.}~\bibnamefont{Harm}},
  \bibinfo{author}{\bibfnamefont{H.-C.} \bibnamefont{N\"agerl}},
  \bibinfo{author}{\bibfnamefont{F.}~\bibnamefont{Ferlaino}},
  \bibinfo{author}{\bibfnamefont{R.}~\bibnamefont{Grimm}},
  \bibinfo{author}{\bibfnamefont{P.~S.} \bibnamefont{Julienne}},
  \bibnamefont{and} \bibinfo{author}{\bibfnamefont{J.~M.}
  \bibnamefont{Hutson}}, \bibinfo{journal}{Phys. Rev. A}
  \textbf{\bibinfo{volume}{87}}, \bibinfo{pages}{032517}
  (\bibinfo{year}{2013}).

\bibitem[{\citenamefont{Vexiau et~al.}(2011)\citenamefont{Vexiau, Bouloufa,
  Aymar, Danzl, Mark, N\"agerl, and Dulieu}}]{vexiau2011}
\bibinfo{author}{\bibfnamefont{R.}~\bibnamefont{Vexiau}},
  \bibinfo{author}{\bibfnamefont{N.}~\bibnamefont{Bouloufa}},
  \bibinfo{author}{\bibfnamefont{M.}~\bibnamefont{Aymar}},
  \bibinfo{author}{\bibfnamefont{J.}~\bibnamefont{Danzl}},
  \bibinfo{author}{\bibfnamefont{M.}~\bibnamefont{Mark}},
  \bibinfo{author}{\bibfnamefont{H.~C.} \bibnamefont{N\"agerl}},
  \bibnamefont{and} \bibinfo{author}{\bibfnamefont{O.}~\bibnamefont{Dulieu}},
  \bibinfo{journal}{Eur. Phys. J. D} \textbf{\bibinfo{volume}{65}},
  \bibinfo{pages}{243} (\bibinfo{year}{2011}).

\end{thebibliography}

\end{document}